\documentclass[secnumarabic,amssymb, nobibnotes, aps]{revtex4}
\usepackage{epstopdf}
\usepackage{amsmath,amssymb,amsthm,amsfonts,mathrsfs,bm,verbatim}
\usepackage{graphicx,subfigure}
\usepackage[utf8]{inputenc}
\usepackage{array}

\newcommand{\bea}{\begin{eqnarray}}
\newcommand{\eea}{\end{eqnarray}}

\newcommand{\bgn}{\begin{align}}
\newcommand{\egn}{\end{align}}

\allowdisplaybreaks[1]

\begin{document}

\title{Radii of spherical photon orbits around Kerr-Newman black holes}

\author{Ying-Xuan Chen}

\affiliation{Guangdong Provincial Key Laboratory of Quantum Engineering and Quantum Materials, School of Physics and Telecommunication Engineering, South China Normal University, Guangzhou 510006, China}
\author{Jia-Hui Huang}
\thanks{huangjh@m.scnu.edu.cn}
\affiliation{{Guangdong Provincial Key Laboratory of Nuclear Science, Institute of quantum matter, South China Normal University, Guangzhou 510006, China}\\
{Guangdong-Hong Kong Joint Laboratory of Quantum Matter, Southern Nuclear Science Computing Center, South China Normal University, Guangzhou 510006, China}\\
{School of Physics and Telecommunication Engineering,South China Normal University, Guangzhou 510006, China}}
\author{Haoxiang Jiang}
\affiliation{Guangdong Provincial Key Laboratory of Quantum Engineering and Quantum Materials, School of Physics and Telecommunication Engineering,South China Normal University,Guangzhou 510006,China}

\begin{abstract}
	The spherical photon orbits around a black hole with constant radii are particular important in astrophysical observations of the black hole. In this paper, the equatorial and non-equatorial spherical photon orbits around Kerr-Newman black holes are studied. The radii of these orbits satisfy a sextic polynomial equation with three parameters, the rotation parameter $u$, charge parameter $w$ and effective inclination angle $v$. For orbits in the polar plane ($v=1$), it is found that there are two positive solutions to the orbit equation, one is inside and the other is outside the event horizon. Particularly, we obtained the analytical expressions for the radii of these two orbits which are functions of $u$ and $w$. For orbits in the equatorial plane ($v=0$) around an extremal Kerr-Newman black hole ($u+w=1$), we find three positive solutions to the equation and provide analytical formulas for the radii of the three orbits. It is also found that a critical value of rotation parameter $u=\frac{1}{4}$ exists, above which only one retrograde orbit exists outside the event horizon and below which two orbits (one prograde and one retrograde) exist outside the event horizon. For orbits in the equatorial plane around a nonextremal Kerr-Newman black hole, there are four or two positive solutions to the corresponding sextic equation, which are shown numerically. The number of solutions depends on the choice of parameters in different regions of $(u,w)$-plane. A critical curve in $(u,w)$-plane which separates these regions is also obtained. On this critical curve, the explicit formulas of three solutions are found. There always exist two equatorial photon orbits outside the event horizon in the nonextremal Kerr-Newman case. For orbits between the above two special planes, there are four or two solutions to the general sextic equation. When the black hole is extremal, it is found that there is a critical inclination angle $v_{cr}$, below which only one orbit exists  and above which two orbits exist outside the event horizon for a rapidly rotating black hole $u>\frac{1}{4}$. At this critical inclination angle, exact formula for the radii is also derived. Finally, a critical surface in parameter space of ($u,w,v$) is shown that separates regions with four or two solutions in the nonextremal black hole case.
\end{abstract}


\maketitle

\section{Introduction}
Black holes are mysterious objects predicted by general relativity. The first black hole merger observation via gravitational waves was reported by LIGO and Virgo in 2016 \cite{LIGOScientific:2016aoc}, which have provided an important window to study the properties of astrophysical black holes \cite{LIGOScientific:2020iuh,LIGOScientific:2020zkf,LIGOScientific:2020ufj}. Significant progress has also been made for observation of black holes via radio waves in recent years.
The first image of the supermassive black hole M87* was provided by Event Horizon Telescope (EHT) in 2019 \cite{EventHorizonTelescope:2019dse,EventHorizonTelescope:2019pgp,EventHorizonTelescope:2019ggy}. In 2022, the image of the supermassive black hole Sagittarius A* in our galaxy center was also released\cite{EventHorizonTelescope:2022xnr}. These images provided us
the appearance of the shadows of the black holes \cite{Cunha:2018acu,Cunha:2017eoe,Gralla:2019xty}.
A shadow is a dark area casted by a black hole for an observer at asymptotic infinity. The radius of the shadow is closely related with the radius of the bound photon orbits around the black hole, which is larger than the event horizon radius \cite{Cvetic:2016bxi,Lu:2019zxb,Ma:2019ybz,Yang:2019zcn,Hod:2020pim,Hod:2013jhd,Liu:2019rib}. In the gravitational wave observation, the frequency and damping time of early-time ringdown signal are related with the properties of unstable photon orbits \cite{Cardoso:2016rao,Cunha:2017eoe}.
Due to the importance in astrophysical observations,  it is worth studying bound null geodesics or photon orbits around a black hole.

For a Schwarzschild black hole with mass $M$, the bound photon orbits are  planar circular orbits, called light rings\cite{Cunha:2017eoe}. The radius of this light ring is $r=3M$ \cite{Darwin1958,Darwin1961,Virbhadra:1999nm,Claudel:2000yi}. All the possible light rings lie in a 2-dimensional spatial hypersurface which is dubbed photon sphere\cite{Virbhadra:1999nm,Claudel:2000yi,Cvetic:2016bxi,Lu:2019zxb,Ma:2019ybz,Yang:2019zcn,Liu:2019rib,Hod:2020pim,Hod:2013jhd}. In fact, for a general spherical static black hole, the bound photon orbits are always light rings  due to the spherical symmetry of the spacetime geometry \cite{Cunha:2017eoe}. However, for axisymmetric and stationary Kerr black holes, there are more kinds of bound photon orbits with constant coordinate $r$ (in Boyer-Lindquist coordinates). In the equatorial plane, two light rings exist outside the black hole horizon. The one with a smaller radius is a prograde orbit moving in the same direction as the black hole and the one with a larger radius is a retrograde orbit\cite{Bardeen:1972fi,Teo2003}. There are also nonplanar spherical photon orbits with constant coordinate radii around a Kerr black hole\cite{Teo2003}. Recently, an interesting topological argument was proposed to prove that at least one light ring exists outside the horizon of a stationary, axisymmetric, asymptotically flat, nonextremal black hole for each rotation sense\cite{Cunha:2020azh}. For extremal rotating black holes, it is found that there exists at least one retrograde light ring which is radially unstable and angularly stable\cite{Guo:2020qwk}. A similar topological proof was extended to black holes in asymptotically flat static de Sitter and anti-de Sitter cases \cite{Wei:2020rbh}. It is also shown that any stationary, axisymmetric, and asymptotically flat spacetime with an ergoregion must have at least one light ring outside the ergoregion \cite{Ghosh:2021txu}.

 Although the analytic formulas between radii of the photon orbits (light rings) in the equatorial and polar planes and the two parameters (mass and angular momentum) of a Kerr black hole have been well studied \cite{Bardeen:1972fi,Teo2003,Hod:2012ax}, the explicit relations between the radii of generic spherical photon orbits and the parameters of the black hole still have been studied in the literature. The radii of the non-equatorial spherical photon orbits around a Kerr black hole was studied in \cite{Yang:2012he,Hod:2012ax}. Approximate formula for the radii of spherical photon orbits was provided in \cite{Hod:2012ax}, which was expressed as a function of the dimensionless angular momentum of the black hole and an effective inclination angle.
  Recently, the exact formulas for the spherical photon orbits which lie between the equatorial and polar planes have been provided. It is found that there exists a critical inclination angle below which there are two null photon orbits outside the black hole event horizon. The radii of the photon orbits are analytically found for the case with the critical inclination angle. For the case with a generic inclination angle, approximate analytical expressions of the radii are provided \cite{Tavlayan:2020cso}.

Kerr-Newman black hole is the most general stationary, axisymmetric, asymptotically flat electrovacuum solution in four-dimensional Einstein-Maxwell theory.
Unlike the Kerr black hole case, the radii of bound photon orbits around a Kerr-Newman black hole are still lack of systematical study. In this work, we consider generic spherical  photon orbits around a Kerr-Newman black hole, focus on the number of photon orbits outside the event horizon and the analytical formulas for the radii of orbits as functions of parameters of the black hole and photon.
The paper is organized as follows. In Section II, we review the null geodesics around a Kerr-Newman black hole and derive the sextic polynomial equation satisfied by the radii of the spherical photon orbits. In Section III, we take three different limits for the Kerr-Newman black hole and check
 our results with previous results in literature. In Section IV, we first study radii of the photon orbits in the polar plane of the black hole. Then, we consider radii of photon orbits in the equatorial plane and summarize our results on Kerr-Newman black holes and known results on other black holes in three tables. Finally, we consider radii of photon orbits between these two special planes. Quantitative and qualitative results are both provided. The last section is devoted to the summary.

\section{Null geodesics around Kerr-Newman black holes}
In the standard Boyer-Lindquist coordinates the metric of a Kerr-Newman black hole with ADM mass $M$, angular momentum $J=Ma$ and electric charge $Q$ (in the $G=c=1$ units) is given by
\bea
ds^2&=&-(1-\frac{2Mr-Q^2}{\Sigma})dt^2+\frac{\Sigma}{\Delta}dr^2+\Sigma d\theta^2
+[(r^2+a^2)\sin^2\theta+\frac{(2Mr-Q^2)a^2\sin^4\theta}{\Sigma}]d\varphi^2\nonumber\\
&&-\frac{2(2Mr-Q^2)a\sin^2\theta}{\Sigma}dtd\varphi,
\eea
where
\begin{gather}
\Sigma=r^2+a^2\cos^2\theta,\\
\Delta=r^2-2Mr+a^2+Q^2.
\end{gather}
$\Delta$ vanishes at
\begin{equation}
	r_\pm={M\pm\sqrt{M^2-a^2-Q^2}},
\end{equation}
which are the radii of outer and inner horizons of the black hole respectively.
 The Lagrangian of a photon is
\begin{equation}
	L=\frac{1}{2}g_{\mu\nu}\dot{x^\mu}\dot{x^\nu},
\end{equation}
where $\dot{x^\mu}=dx^\mu/d\lambda$ is the 4-velocity of the photon. $\lambda$ is an affine parameter along the photon orbit.
There are two Killing vectors $\xi_{(t)}=\partial/\partial t$ and $\xi_{(\varphi)}=\partial/\partial\varphi$ for the Kerr-Newman geometry which provide two conserved quantities along the light geodesic. One is conserved energy $E$ of the photon and the other one is conserved azimuthal angular momentum $L_z$.  There is a third constant of motion $\kappa$, called Carter constant \cite{Carter:1968rr}, which is related with a symmetric rank two Killing tensor and determines the behavior of the photon's motion in the $\theta$-direction\cite{Teo2003}. $\kappa=0$ characterizes the equatorial motion of photon \cite{Hod:2012ax,Tavlayan:2020cso}. A relevant property of the Carter constant is $\kappa\geqslant 0$\cite{Hod:2012ax,Tavlayan:2020cso}.

Using the above mentioned three constants of motion, the geodesic motion of the photon around the Kerr-Newman black hole is governed by the following set of first-order differential equations (we define a new variable $o\equiv\cos\theta$ in place of the coordinate $\theta$)
\bea
\Sigma\dot{r}&=&\pm\sqrt{R(r)},\label{r-eq}\\
\Sigma\dot{o}&=&\pm\sqrt{V(o)},\label{o-eq}\\
\Sigma\dot{\varphi}&=&\frac{L_z}{1-o^2}+\frac{a}{\Delta}[(2Mr-Q^2)E-a L_z],\\
\Sigma\dot{t}&=&\frac{1}{\Delta}[(r^2+a^2)^2E-a L_z(2Mr-Q^2)]-a^2E(1-o^2),
\eea
where
\begin{align}
	R(r)=&E^2r^4+(a^2E^2-\kappa-L_z^2)r^2+2M[\kappa+(L_z-aE)^2]r-a^2\kappa-Q^2[\kappa+(L_z-aE)^2],\label{10}\\
	V(o)=&-a^2E^2o^4+[a^2E^2-L_z^2-\kappa]o^2+\kappa.
\end{align}
 Equations \eqref{r-eq}\eqref{o-eq} determine the orbital motions in $r$ and $\theta$ directions.

We are interested in the spherical photon orbits with constant radius $r$.  They are determined by the following two conditions\cite{Bardeen:1972fi,Teo2003,Hod:2012ax,Tavlayan:2020cso}:
\begin{equation}
R(r)=0,~~\frac{dR(r)}{dr}=0.
\end{equation}
Plugging equation \eqref{10} into the above equations, we can obtain the following two equations,
\bea
\frac{L_z}{E}&=&-\frac{r^3-3Mr^2+a^2r+a^2M+2Q^2r}{a(r-M)}\label{13},\\
\frac{\kappa}{E^2}&=&-\frac{r^2(r^4-6Mr^3+9M^2r^2-4a^2Mr+4Q^2r^2-12MQ^2r+4Q^4+4a^2Q^2)}{a^2(r-M)^2}\label{14}.
\eea
To describe the deviation of an orbit from the equatorial plane of the black hole, we can define the following effective inclination angle \cite{Hod:2012ax,Tavlayan:2020cso}
\begin{equation}
	\cos i\equiv\frac{L_z}{\sqrt{L^2_z+\kappa}}.
\end{equation}
It is easy to see that the effective inclination angle is a constant of motion. The orbits on the equatorial plane are characterized by $\cos i=\pm 1$ (i.e. $\kappa=0$) and orbits on the polar plane are characterized by $\cos i=0$ (i.e. $L_z=0$).

To simplify the analysis of the radii of generic spherical photon orbits, we define the following dimensionless parameters
\begin{gather}
x\equiv\frac{r}{M},~~u\equiv\frac{a^2}{M^2},~~w\equiv\frac{Q^2}{M^2},~~v\equiv \sin^2i.
\end{gather}
The above parameters should satisfy the following physical constraints:
\bea\label{fanwei}
x\ge 0,~~0\le u\leq1,~~0\le w\leq1,~~0\le u+w\leq1,~~0\le v\leq1.
\eea
Eliminating the photon's energy $E$ in equations \eqref{13}\eqref{14}, the following sextic polynomial equation can be obtained
\begin{gather}
	P_6(x)=x^6-6x^5+(9+2uv+4w)x^4-(4u+12w)x^3-(6uv-u^2v-4w^2-4uw)x^2+(2u^2v+4wuv)x+u^2v=0.\label{16}
\end{gather}
The positive real roots of this equation provide us the radii of generic spherical photon orbits.

\section{Three special cases}
In this section we consider the solutions to the sextic equation \eqref{16} for three special cases, which correspond to different limiting cases of the Kerr-Newman black hole.

When the rotation parameter $u=0$ and charge parameter $w=0$, the Kerr-Newman black hole becomes a Schwarzschild black hole. In this case, the sextic polynomial equation \eqref{16} reduces to (ignoring the zero root)
\begin{gather}
	x-3=0.
\end{gather}
So there is only one spherical photon orbit with radius $x=3 (r=3M)$ around a Schwarzschild black hole. The event horizon of the Schwarzschild black hole is located at $r=2M$, so the radius of the photon orbit is larger than that of the event horizon.

When we take a vanishing rotation parameter $u=0$ and nonzero charge parameter $w\neq 0$, the Kerr-Newman black hole becomes a Reissner-Nordstrom black hole. In this case, the sextic polynomial equation \eqref{16} reduces to (ignoring the zero root)
\begin{gather}
x^2-3x+2w=0.
\end{gather}
Considering the range of parameter $w$ in \eqref{fanwei}, we find that there are two real roots for the above equation, which are
\begin{gather}
	x_{1}=\frac{1}{2}(3-\sqrt{9-8w}),~~
	x_{2}=\frac{1}{2}(3+\sqrt{9-8w}).
\end{gather}
The inner and outer horizons of Reissner-Nordstrom black hole are located at $x_\pm=1\pm\sqrt{1-w}$. It is easy to see that the orbit with smaller radius lies between the two black hole horizons, i.e.
\bea
x_-\le x_{1}\le x_+,
\eea
which are saturated simultaneously in the extremal Reissner-Nordstrom case ($w=1$).

The photon orbit with a larger radius, $x_{2}$, always lies outside the black hole event horizon. The range
of the dimensionless radius $x_{2}$ is
\bea
2\le r_{2}\le 3.
\eea
The radius of this photon orbit monotonically decreases as the amount of black hole charge increases.
For the mostly charged extremal Reissner-Nordstrom black hole, $x_{2}=2$ and the event horizon is located at $x=1$.

It is worth mentioning that due to the spherical symmetry of the Schwarzschild and Reissner-Nordstrom black holes, the orbit radii in this two cases are independent of the effective inclination parameter $v$.

When we take a non-vanishing rotation parameter $u$ and zero charge parameter, $w=0$, the Kerr-Newman black hole becomes a Kerr black hole. In this case, the sextic polynomial equation \eqref{16} reduces to
\bea
	x^6-6x^5+(9+2uv)x^4-4ux^3-(6uv-u^2v)x^2+2u^2vx+u^2v=0.
\eea
For equatorial orbits, the effective inclination parameter $v=0$. The above equation further reduces to
\bea\label{kerreq}
x^3-6x^2+9x-4u=0.
\eea
A detailed study of the above equation and the more generic spherical photon orbits was in \cite{Tavlayan:2020cso}, including equatorial orbits, polar orbits and the case where orbits lie between these two extremes.

\section{Photon orbits around Kerr-Newman black holes}
\subsection{Polar orbits}
\begin{figure}[htb]
	\includegraphics[width=8cm]{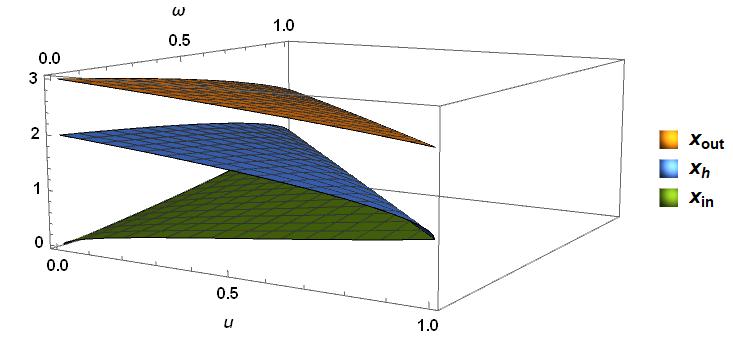}
	\caption{Two positive solutions $x_{in}$,$x_{out}$ in polar plane and the event horizon radius $x_h$ are plotted as functions of $u$ and $w$.}
	\label{f4}
\end{figure}

In this subsection, we consider the photon orbits in the polar plane of a Kerr-Newman black hole. In this case, the effective inclination parameter $v=1 (L_z=0)$.
The equation satisfied by the radii of polar orbits is (ignoring the zero root)
\begin{gather}
	x^3-3x^2+(2w+u)x+u=0.
\end{gather}
There are three real roots for the above equation, one of which is negative and the other two are positive. The explicit analytical expressions of these roots are
\bea
x_{neg}&=&1-2\sqrt{1-\frac{u+2w}{3}}\sin\left(\frac{\pi}{6}+\frac{1}{3}\arccos\left(\frac{1-u-w}{[1-(u+2w)/3]^{3/2}}\right)\right),\\
x_{in}&=&1-2\sqrt{1-\frac{u+2w}{3}}\sin\left(\frac{\pi}{6}-\frac{1}{3}\arccos\left(\frac{1-u-w}{[1-(u+2w)/3]^{3/2}}\right)\right),\\
x_{out}&=&1+2\sqrt{1-\frac{u+2w}{3}}\cos\left(\frac{1}{3}\arccos\left(\frac{1-u-w}{[1-(u+2w)/3]^{3/2}}\right)\right).
\eea
The negative root $x_{neg}$ is unphysical. For different rotation and charge parameters, the range of the inner orbit $ x_{in}$ is $0 \le x_{in}\le 1$ and
the range of the outer orbit $ x_{out}$ is $1\le x_{out}\le 3$. In Fig.\ref{f4}, we plot all three roots and the event horizon radius as functions of the rotation parameter $u$ and charge parameter $w$. We can see that inner orbit lies in the interior of the black hole horizon and the outer orbit always lies outside the event horizon.
So in the polar plane of a Kerr-Newman black hole, we have only one spherical photon orbit outside the event horizon with radius
\bea
r_{polar}=M+2M\sqrt{1-\frac{u+2w}{3}}\cos\left(\frac{1}{3}\arccos\left(\frac{1-u-w}{[1-(u+2w)/3]^{3/2}}\right)\right).
\eea
When taking the charge parameter $w$ to zero, one can check that the above result is the same as that obtained in the Kerr black hole case \cite{Tavlayan:2020cso}.

\begin{figure}[htb]
	\includegraphics[width=6cm]{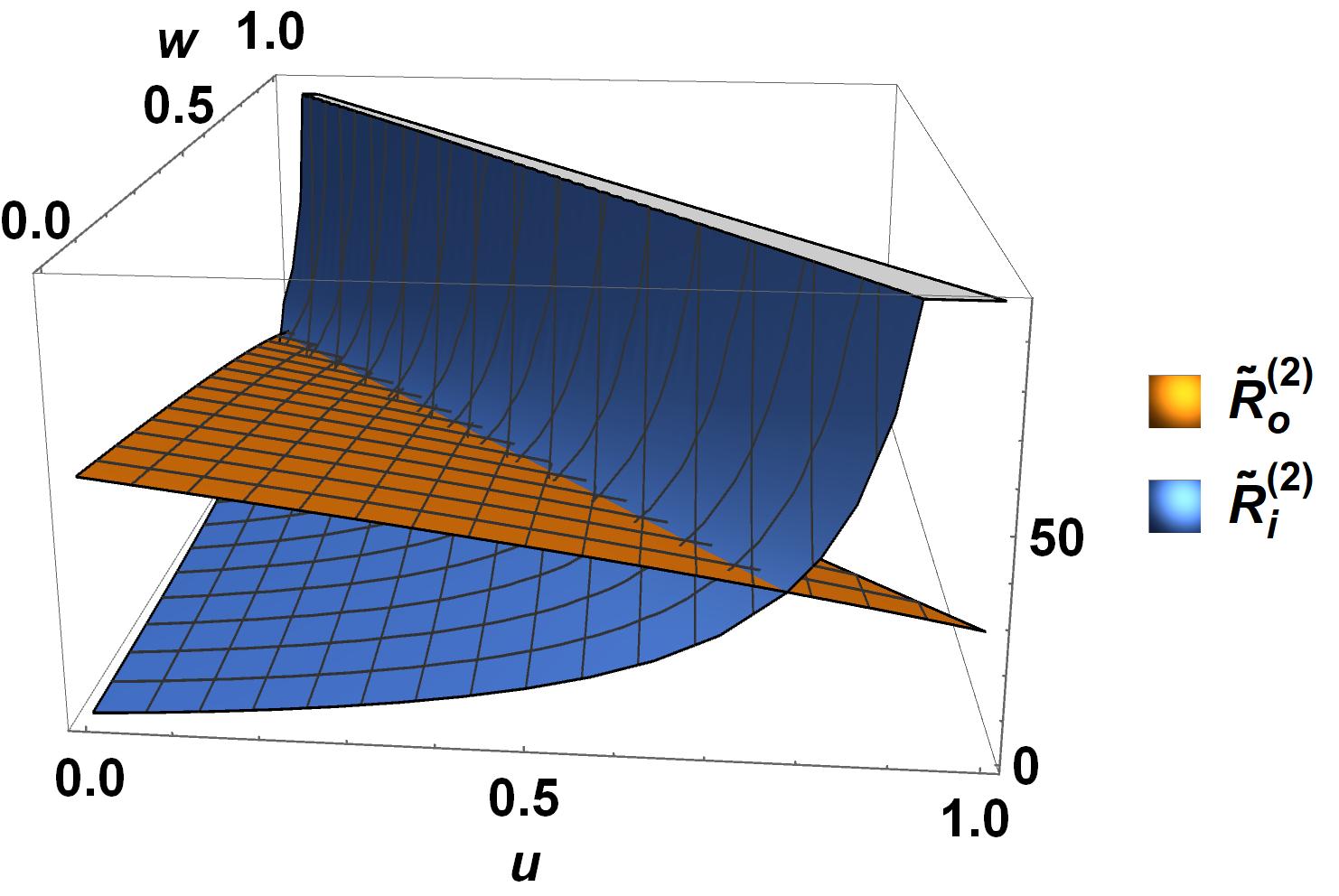}
	\caption{$\tilde{R}^{(2)}_o,~\tilde{R}^{(2)}_i$ are positive as functions of $u$ and $w$. }
	\label{f5}
\end{figure}

Then we analyze the radial stability of the outer and inner polar orbits by considering $\tilde{R}^{(2)}_o\equiv\frac{1}{M^4E^2}\frac{d^2R_o}{dx^2}$ and $\tilde{R}^{(2)}_i\equiv\frac{1}{M^4E^2}\frac{d^2R_i}{dx^2}$ respectively.
They are plotted as functions of parameters $u$ and $w$ in Fig.\ref{f5}. We can see that these two functions are always positive. So the two solutions are both unstable under radial perturbation.

\subsection{Equatorial orbits}
For photon orbits in the equatorial plane, the effective inclination parameter $v=0$. From equation \eqref{16}, we can obtain the equation satisfied by the radii of equatorial orbits, which is (ignoring the zero root)
\bea\label{kneq}
x^4-6x^3+(9+4w)x^2-(4u+12w)x+4w(w+u)=0.
\eea

Let's first consider a simpler case, the equatorial orbits around an extremal Kerr-Newman black hole. The rotation parameter $u$ and charge parameter $w$ of the black hole satisfy $u+w=1$.
The above equation can be further reduced to the following form
\bea
(x-1)^2(x^2-4x+4w)=0.
\eea
Since $0\le w\le 1$, three real roots of the above equation are
\bea
x_1=1,~~x_2=2(1-\sqrt{1-w}),~~x_3=2(1+\sqrt{1-w}).
\eea
The double root $x=1$ has the same radial coordinate as the event horizon. For a highly charged (slowly rotating) extremal Kerr-Newman black hole with $\frac{Q}{M}>\frac{\sqrt{3}}{2}$ or $w>\frac{3}{4}$($u<\frac{1}{4}$), there are two photon orbits outside the event horizon with radii $x_2=2(1-\sqrt{1-w})$ and $x_3=2(1+\sqrt{1-w})$. The orbit with a smaller radius $x_2$ is prograde and the orbit with a larger radius $x_3$ is retrograde.  For a lowly charged (rapidly rotating) extremal Kerr-Newman black hole with $\frac{Q}{M}\le\frac{\sqrt{3}}{2}$ or $w\le\frac{3}{4}$($u\ge\frac{1}{4}$), there is only one retrograde photon orbit outside the event horizon with radius $x_3=2(1+\sqrt{1-w})$.

Now let's consider the equatorial orbits around a nonextremal Kerr-Newman black hole. Due to the nonzero charge of the rotating black hole, the generic equation \eqref{kneq} satisfied by the orbit radii is a quartic equation which is more complicate than the cubic equation \eqref{kerreq} of the Kerr black hole case. In order to simplify the generic quartic equation, we change the variable $x$ to $t\equiv x-3/2$, then the generic equation \eqref{kneq} can be rewritten as
\bea\label{kneqt}
\left(t^2-\frac{9-8w}{4}\right)^2=4u\left(t-\frac{2w-3}{2}\right).
\eea
When parameters $u$ and $w$ are in their physical regime \eqref{fanwei}, we find that there are four or two real roots for the above equation when the choices of ($u,w$) are in  different regions in ($u,w$)-plane.

\begin{figure}[h]
	\includegraphics[width=6cm]{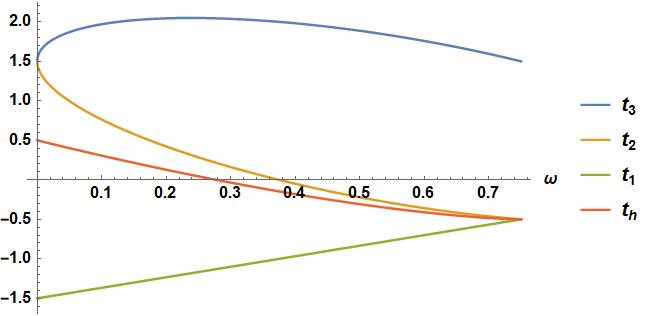}
	\caption{In critical case, equatorial orbit solutions $t_1, t_2, t_3$ and event horizon $t_h$ are plotted as functions of charge parameter $w$.}{\label{eqcr}}
\end{figure}

A critical curve in the $(u,w)$-plane is found that separates the different regions in ($u,w$)-plane.
This critical curve is described by the following equation
\bea
u_{cr}=\frac{w}{27}(8w-9)^2,
\eea
where $w<3/4$ because of the constraint $u+w<1$. When $u$ and $w$ are on the critical curve, the generic quartic equation \eqref{kneqt} becomes into
\bea
(6t-8w+9)^2(12 t^2+ 4(8w-9)t +27-96 w+64 w^2)=0.
\eea
The real roots of the above equation are
\bea
t_1=\frac{1}{6}(8w-9),~t_2= \frac{1}{6}(9 - 8w - 4 \sqrt{9w-8w^2}),~t_3=\frac{1}{6}(9 - 8w + 4 \sqrt{9w-8w^2}).
\eea
A double root $t_1$ appears. In this critical case, the location of the event horizon of the Kerr-Newman black hole in coordinate $t$ is
\bea
t_h=-\frac{1}{2}+\left(1-\frac{4}{3}w \right)^{3/2}.
\eea
The orbit $t_1$ is inside the event horizon and the orbits $t_2$ and $t_3$ are outside the event horizon. This is shown in Fig.\ref{eqcr}.

\begin{figure}[h]
	\includegraphics[width=6cm]{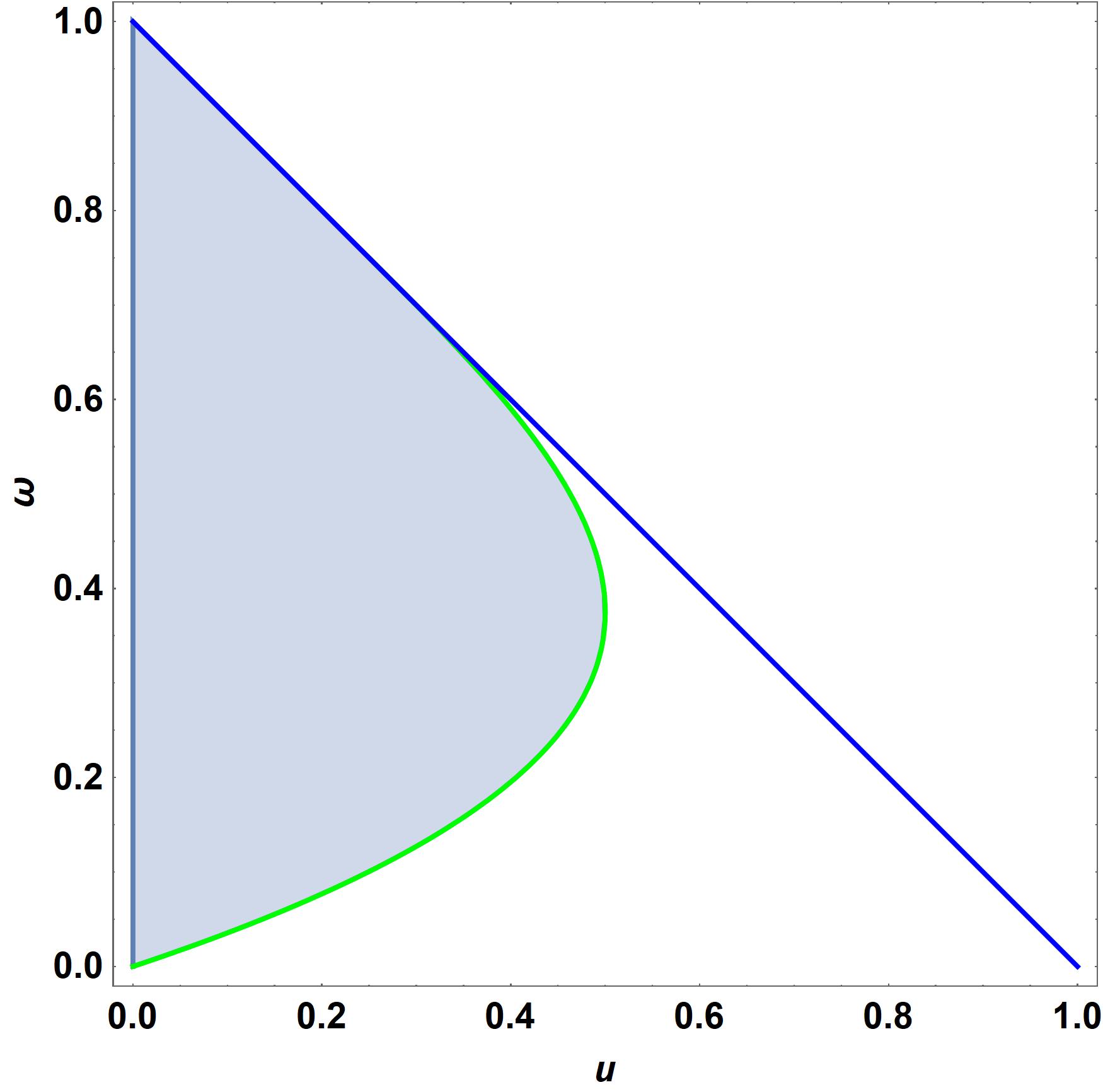}
	\caption{The critical curve (Green) and the two parts of the physical $(u,w)$-plane (lower left triangle). When $(u,w)$ is in the shaded part, two roots exist
for the equatorial orbit equation. When $(u,w)$ is in the other part, four roots exist
for the equatorial orbit equation.  }{\label{uw}}
\end{figure}

This critical curve divides the physical region (lower left triangle) in the $(u,w)$-plane into two parts. When $(u,w)$ belongs to the shaded part, there are two real roots for the orbit equation. When $(u,w)$ belongs to the unshaded part, there are four real roots for the orbit equation. This is shown in Fig.\ref{uw} and the critical curve is drawn in green. The intersection point of this critical curve and the line $u+w=1$ is $(u,w)=(\frac{1}{4},\frac{3}{4})$.

\begin{figure}[h]
	\includegraphics[width=6cm]{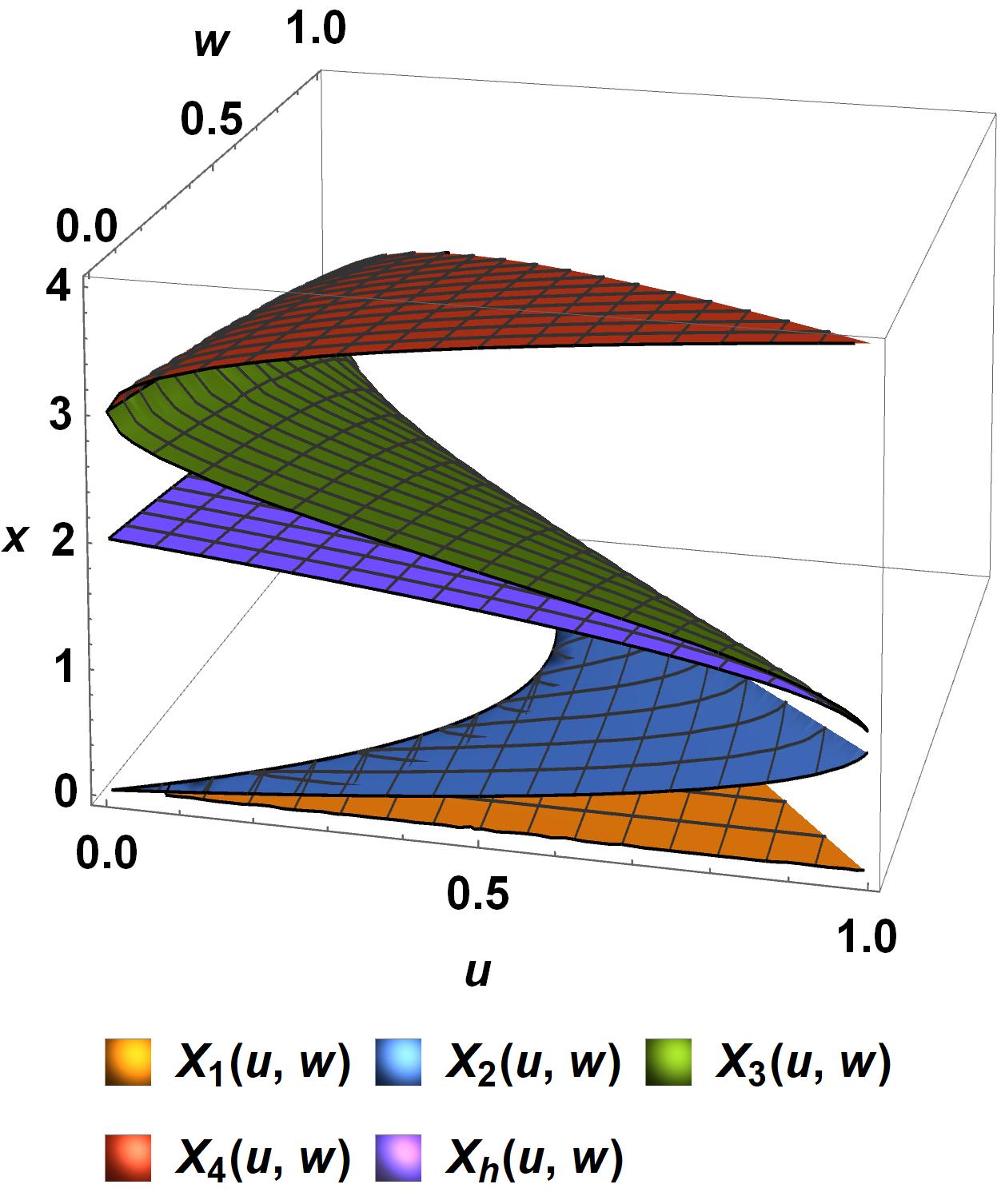}
	\caption{The roots of the equation of equatorial orbits $x_1,x_2,x_3,x_4$ and the black hole event horizon $x_h$  are shown as functions of $(u,w)$. There are always two orbits outside the event horizons.}{\label{f1}}
\end{figure}

Although there exit analytic solutions for the generic quartic equation \eqref{kneq}, their expressions are cumbersome and pointless. We just numerically show these solutions as functions of the rotation parameter $u$ and the charge parameter $w$ in Fig.\ref{f1} for illustration. From this figure, we can indeed see that there are only two real roots in one part of the $(u,w)$-plane while in the other part there are four real roots. The boundary of these two parts is consistent with the critical curve given before. In this figure, we also show the event horizon of the Kerr-Newman black hole, $x_h=1+\sqrt{1-u-w}$. We can see that there always exist two equatorial photon orbits outside the event horizon in the nonextremal case, which is consistent with the general topological proofs on the existence of photon orbits \cite{Cunha:2020azh,Guo:2020qwk}. These radii are closely related with the critical impact parameter of equatorial light bending around the Kerr-Newman black hole\cite{Hsiao:2019ohy,Hsieh:2021scb}.

\begin{figure}[h]
	\includegraphics[width=6cm]{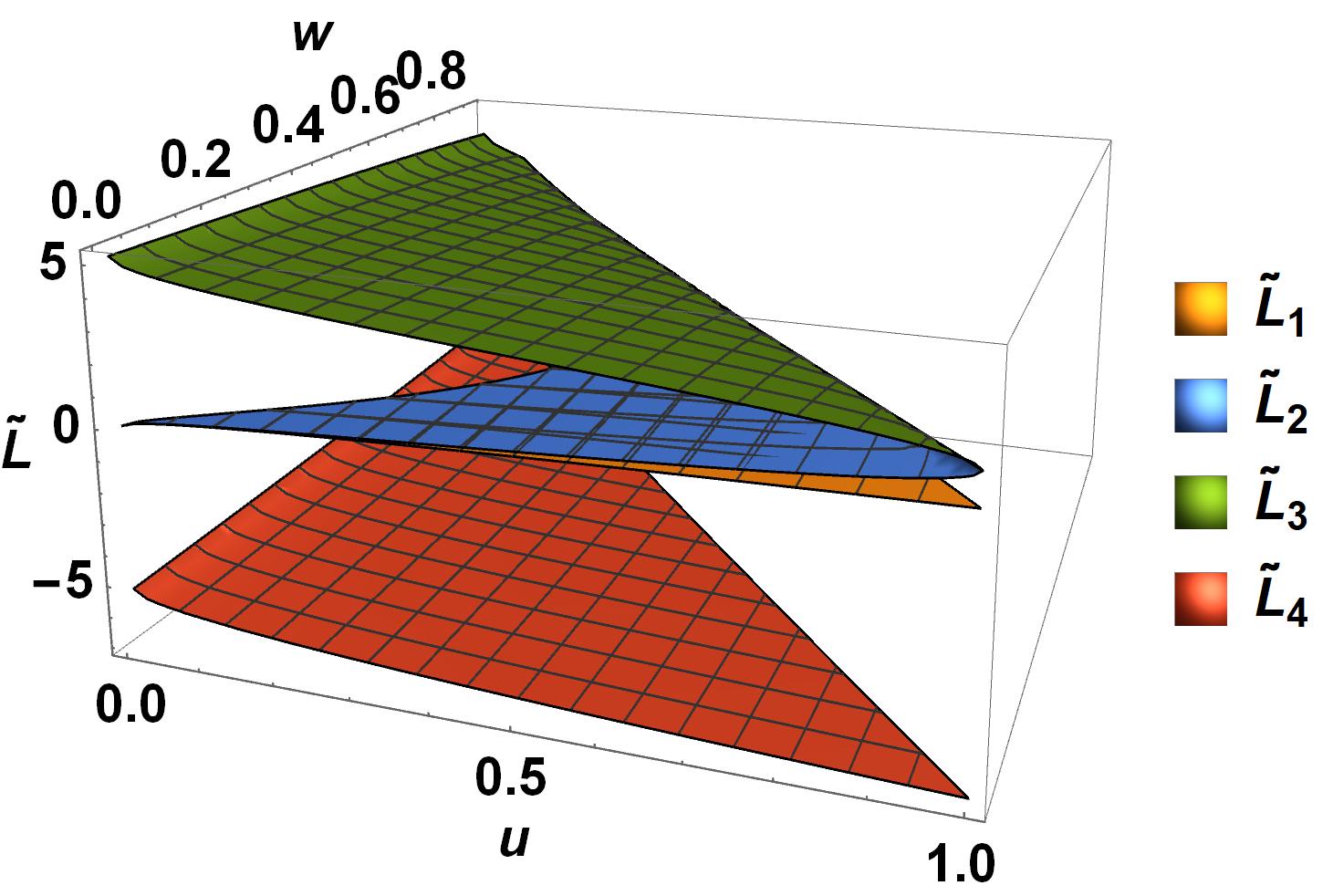}
	\caption{The dimensionless $z$-component of the orbital angular momentum per energy $\tilde{L}$ for the orbits.  $\tilde{L}_1$,$\tilde{L}_2$,$\tilde{L}_3$,$\tilde{L}_4$ correspond to the orbits $x_1,x_2,x_3,x_4$ respectively. $\tilde{L}_3>0$, the inner orbit $x_3$ is prograde. $\tilde{L}_4<0$, the outer orbit $x_4$ is retrograde.}{\label{lz}}
\end{figure}

\begin{figure}[h]
	\includegraphics[width=6cm]{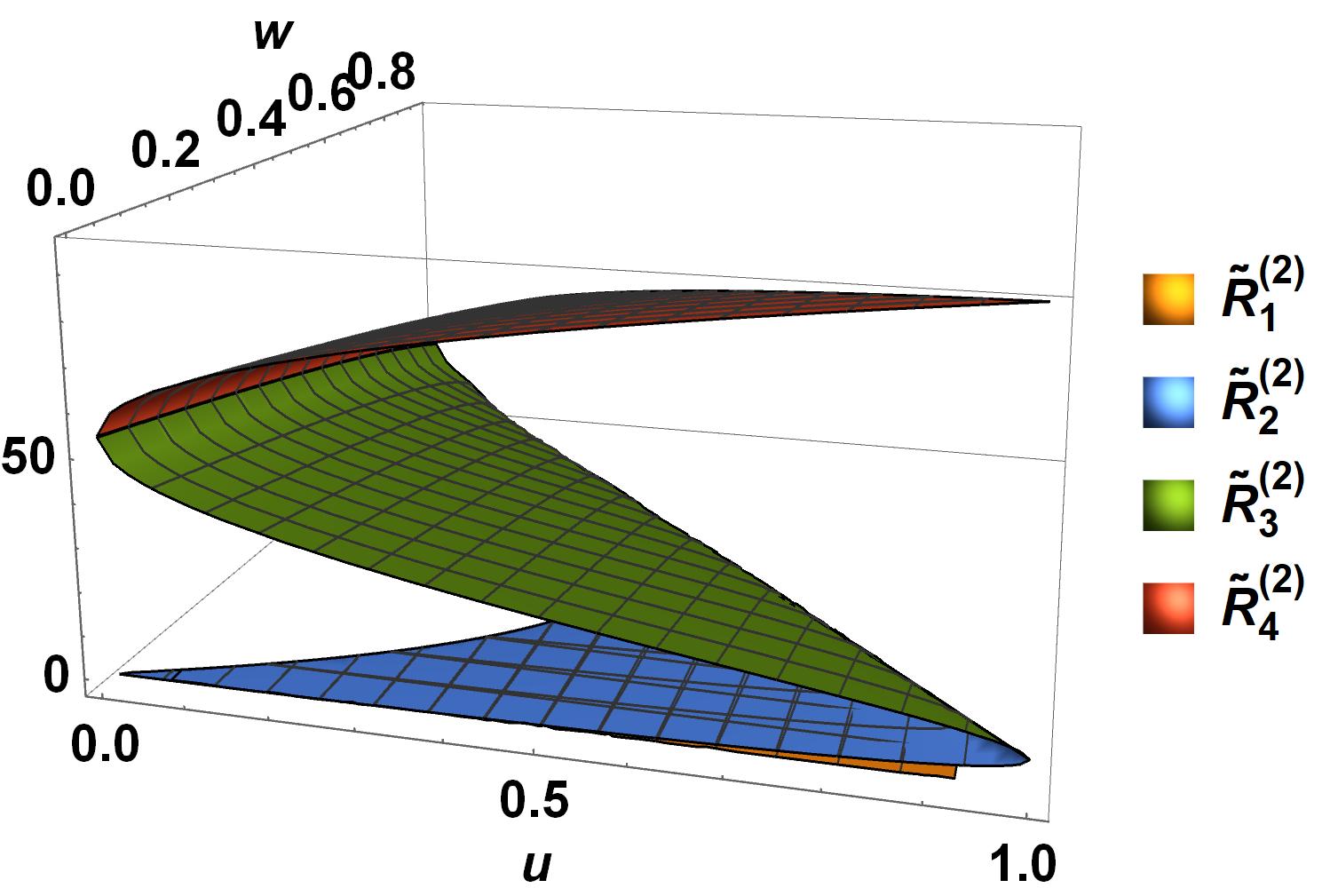}
	\caption{The second derivatives $\tilde{R}^{(2)}_i$ as functions of $u$ and $w$ for equatorial solutions $x_i$. The solutions are radially unstable since $\tilde{R}^{(2)}_i>0$.}{\label{FIG.3}}
\end{figure}

After obtaining the location  of an equatorial orbit $x_i(u,w)$, we study the $z$-component of the orbital angular momentum $L_z$ of a photon in this orbit by using equation \eqref{13}. When $L_z$ is positive, the photon rotates in the same direction with the black hole and this orbit is prograde. When $L_z$ is negative, the photon rotates against the direction of the black hole rotation and this orbit is retrograde. In Fig.\ref{lz}, we show the dimensionless $z$-component of the angular momentum per energy, $\tilde{L}=\frac{L_z}{ME}$, for each equatorial orbit. We can see that for the two orbits outside the horizon, the outer one is retrograde and the inner one is prograde.
 By calculating the second derivative of $\tilde{R}(x)\equiv\frac{R(r)}{M^4 E^2}$ at an orbit, we can obtain its radial stability. Defining $\tilde{R}^{(2)}_i=\frac{d^2\tilde{R}(x)}{dx^2}|_{x=x_i}$, we show $\tilde{R}^{(2)}_i$ in Fig.\ref{FIG.3} as functions of $u$ and $w$. We can see that $\tilde{R}^{(2)}_i>0$, which means the orbits are radially unstable.


 Based on our results in this subsection and results in previous literatures, we completely know some aspects on the equatorial photon orbits around Kerr-Newman black holes and the various limiting black holes. In Table.\ref{tab1}, Table.\ref{tab2} and Table.\ref{tab3},
we summarize the positive solutions of the equatorial orbit equation \eqref{kneq}, $n_s$, and the equatorial photon orbits outside event horizon, $n_{o}$ for different black holes. For several cases where the analytic expressions are very complicate, we just indicate the numbers of the solutions and orbits. In the tables, S, RN, eRN, K, eK, KN and eKN mean Schwarzschild, nonextremal Reissner-Nordstrom, extremal Reissner-Nordstrom, nonextremal Kerr, extremal Kerr, nonextremal Kerr-Newman and extremal Kerr-Newman black holes respectively.
\begin{table}[htbp]
\centering\caption{Equatorial photon orbits for general KN black holes.}\label{tab1}
\begin{tabular}{|c|c|c|c|c|c|}
  \hline
  BHs & $w=1,u=0$(eRN) & $0<w<1,u=0 $(RN) & $w=u=0$(S) & $w=0,0<u<1$(K) &$w=0,u=1 $(eK) \\ \hline
  $n_s$ &$x_1=1,x_2=2$  &$x_\pm=\frac{3\pm\sqrt{9-8w}}{2}$  &$x=3$  &$x_\pm=2+2\cos(\frac{2\arccos(\pm\sqrt{u})}{3}), x_3=4\sin^2(\frac{\arccos\sqrt{u}}{3})$  & $x_1=1,x_2=4$  \\ \hline
  $n_{o}$ & $x_2=2$ & $x_+=\frac{3+\sqrt{9-8w}}{2}$ & $x=3$  & $x_\pm=2+2\cos(\frac{2\arccos(\pm\sqrt{u})}{3})$ & $x=4$ \\
  \hline
\end{tabular}
\end{table}
\begin{table}[htbp]
\centering\caption{Equatorial photon orbits for general KN black holes.$*$A pure integer indicates the number of solutions.}\label{tab2}
\begin{tabular}{|c|c|c|c|c|}
  \hline
  BHs & $u+w=1,0<w<\frac{3}{4}$(eKN) & $u=\frac{1}{4}, w=\frac{3}{4}$(eKN) & $u+w=1,\frac{3}{4}<w<1$(eKN) & $\frac{3}{4}\le w<1,0<u<1-w$(KN)  \\ \hline
  $n_s$ &$x_1=1,x_\pm=2(1\pm\sqrt{1-w})$  &$x_1=1,x_2=3$  &$x=1,x_\pm=2(1\pm\sqrt{1-w})$  &2*   \\ \hline
  $n_{o}$ & $x_+=2(1+\sqrt{1-w})$ & $x=3$ & $x_\pm=2(1\pm\sqrt{1-w}) $  & 2 \\
  \hline
\end{tabular}
\end{table}
\begin{table}[htbp]
\centering\caption{Equatorial photon orbits for general KN black holes.$*$A pure integer indicates the number of solutions.}\label{tab3}
\begin{tabular}{|c|c|c|c|c|}
  \hline
  BHs & $0<w<\frac{3}{4},0<u<u_{cr} $(KN) & $0<w<\frac{3}{4},u=u_{cr} $(KN)  & $0<w<\frac{3}{4},u_{cr}<u<1-w $(KN)  \\ \hline
  $n_s$ &2*  &$x_1=\frac{4}{3}w,x_\pm= \frac{1}{6}(18 - 8w\pm 4 \sqrt{9w-8w^2})$ &4   \\ \hline
  $n_{o}$ & 2& $x_\pm= \frac{1}{6}(18 - 8w\pm 4 \sqrt{9w-8w^2})$& 2   \\
  \hline
\end{tabular}
\end{table}

\subsection{Photon orbits for general cases}
In this subsection, we consider the most general case for spherical photon orbits around a Kerr-Newman black hole, where three parameters $u,w,v$ are all nonzero.
The radii of these orbits are the solutions of the sextic equation \eqref{16}. Since there are four sign changes in this polynomial equation, according to the Descartes' rule of signs, we know that the possible numbers of positive roots of this sextic equation may be 4, 2 or 0.
It is known that there is no analytical solution for a general algebraic equation higher than quartic. Numerical or approximate analytical methods should be used to obtain the radii of the orbits. But in certain special cases, we could get analytical solutions for the sextic equation.

\begin{figure}[htb]
	\includegraphics[width=8cm]{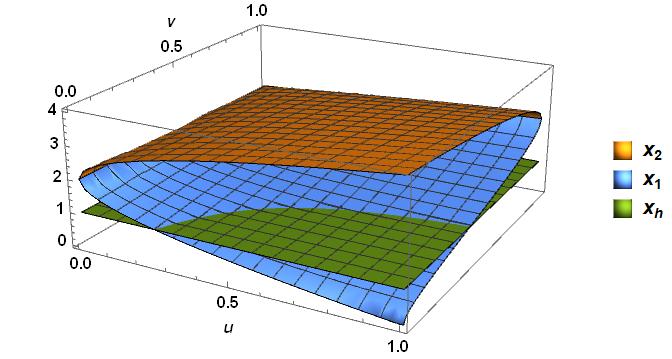}
	\caption{In the eKN case, two positive real roots $x_1, x_2$ and the event horizon $x_h$ are plotted as functions of rotation parameter $u$ and effective inclination angle $v$. }
	\label{eknvf}
\end{figure}

Let's first consider the extremal Kerr-Newman black hole case, where $u+w=1,0<u<1,0<w<1$. We substitute $w$ for $1-u$ and rewrite the sextic equation \eqref{16} as follows
\bea\label{eknoeq}
(x-1)^2P_4(x)=0,
\eea
where
\bea
P_4(x)=x^4-4x^3+(4 - 4 u + 2 u v)x^2+4u v x+u^2v.
\eea
A double root $x=1$ appears and has the same radial coordinate as the event horizon. According to the Descartes' rule of signs, we know there are at most 2 positive real roots for equation $P_4(x)=0$. With numerical method, we confirm that $P_4(x)=0$ has exactly two different positive real roots $x_1,x_2$. The analytical expressions of these two roots are cumbersome and we just show them in Fig.\ref{eknvf}. In this figure, we also show the location of the event horizon $x_h=1$. We can see that the values of the roots satisfy $0<x_1<x_2<4$.
When we just consider the photon orbits outside the event horizon, it is obvious that there are two photon orbits for one region of the ($u,v$)-plane and one orbit for the other region.

Explicitly, we find that for a slowly rotating extremal KN black hole, i.e. $0<u<\frac{1}{4}$, there always exist two photon orbits outside the event horizon for any effective inclination angle ($0<v<1$). But, for a rapidly rotating extremal KN black hole, i.e. $\frac{1}{4}<u<1$, there exists a critical inclination angle $v_{cr}$. When $v>v_{cr}$, there are two photon orbits outside the event horizon ($x_1$ is prograde, $x_2$ is retrograde) and when $v<v_{cr}$, there is only one retrograde photon orbit outside the event horizon. It is found that this critical effective inclination angle $v_{cr}$ is determined by the following equation
\bea
v_{cr}=\frac{-1 + 4 u}{u (u+6)},
\eea
where $\frac{1}{4}<u<1$. When $(u,v)$ is on the above curve, equation \eqref{eknoeq} can be rewritten as
\bea
(x-1)^3\left((6 + u) x^3-(18 + 3 u) x^2 + (4 - 15 u - 4 u^2) x+ u - 4 u^2 \right)=0.
\eea
 $x=1$ becomes a triple root. Because $u>\frac{1}{4}$, we find that there is only one real root for the following equation
 \bea
 (6 + u) x^3-(18 + 3 u) x^2 + (4 - 15 u - 4 u^2) x+ u - 4 u^2 =0,
 \eea
 which is
 \bea
 x_{cr}=1+T^{1/3}+\frac{2(2 u+1)(u+7)}{3(6 + u)}T^{-1/3},\\
 T=\frac{4(1+u)^2}{6+u}+\sqrt{\frac{-8 (1 + u)^3 (19 - 84 u + 30 u^2 + 8 u^3)}{27 (6 + u)^3}}.
 \eea

\begin{figure}[htb]
	\includegraphics[width=6cm]{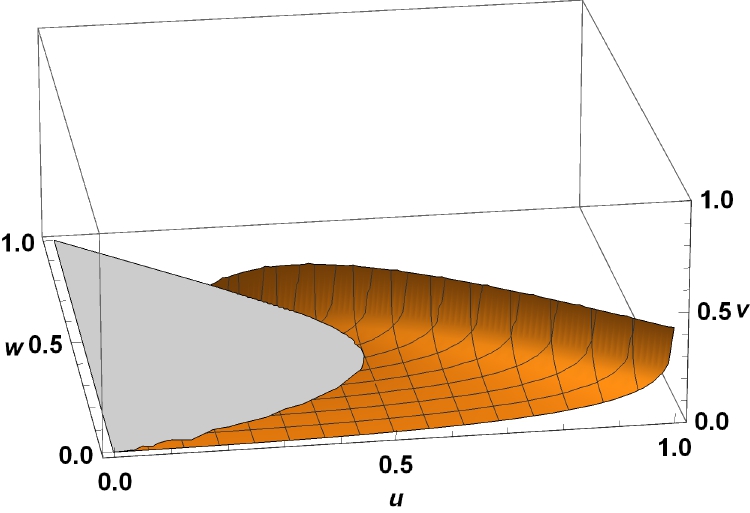}
	\caption{The critical inclination angle $v_{s}$ is plotted as a function of the rotation parameter $u$ and the charge parameter $w$.}
	\label{f6}
\end{figure}

Now let's consider the nonextremal Kerr-Newman case. We know that the possible numbers of the solutions are four, two and zero.
It is hard to get analytical solutions for the sextic equation in this case. But, given the three parameters ($u,v,w$), we can obtain the solutions with numerical method. Here we mainly focus on the qualitative property of the solutions.
The solutions depend on three parameters ($u,v,w$). One can imagine there are different numbers of solutions when ($u,v,w$) is in different regions of the parameter space. There should exist a critical curved surface that separates the regions.
This surface is likely to be determined by the condition that  the sextic equation has multiple roots. The appearance of multiple roots is equivalent to the vanishing of the discriminant of the sextic polynomial equation, which is defined as
\begin{gather}
	D=(-1)^\frac{n(n-1)}{2}\mathcal R(P_6(x),P'_6(x)),
\end{gather}
where $P'_6(x)$ is the derivative of $P(x)$ and $\mathcal R$ is the eliminant of $P_6(x)$ and $P'_6(x)$. From $D=0$, we can get the analytic expression of the critical surface, $v=v_s(u,w)$. However, the explicit expression is rather cumbersome, so we only show it in FIG.\ref{f6}. There are four roots when the three parameters are below the critical surface, while there are only two roots above. This is consistent with our previous results that there are only two roots in the polar plane but four or two roots in the equatorial plane.

\section{summary}
In this paper, we analytically and numerically study the spherical photon orbits around general Kerr-Newman black holes. Starting from the radial null geodesic equation, we derive the sextic equation satisfied by the radii of the spherical photon orbits, which depends on three parameters, the rotation parameter $u$, the charge parameter $w$ and effective inclination angle $v$. After checking three limiting cases, we mainly study the number of photon orbits and the exact analytical expressions of the radii around Kerr-Newman black holes. The main results of this paper are:

 (1) In the polar plane ($v=1$), it is found that there are two positive solutions to the corresponding orbit equation. But only one spherical photon orbit outside the event horizon. The exact analytical formula for these radii are provided, which reduce to the known formula of the Kerr case \cite{Hod:2012ax,Tavlayan:2020cso} when taking zero charge limit. The radius of the orbit outside the event horizon is the same as the result of a recently work \cite{Wang:2022ouq}.

 (2) In the equatorial plane($v=0$), there are four or two solutions to the corresponding orbit equation, depending on the choice of the rotation parameter $u$ and charge parameter $w$. A critical curve, $u_{cr}=\frac{w}{27}(8w-9)^2$, is found in the $(u,w)$ plane, that separates two regions corresponding to different number of solutions. When $(u,w)$ is on this curve, the exact analytical formula for three radii are found (one solution is a double root). There always exist two equatorial photon orbits outside the event horizon in this nonextremal Kerr-Newman black hole case, which is consistent with the general topological arguments on the existence of photon orbits \cite{Cunha:2020azh,Guo:2020qwk}. We summarize the results on equatorial photon orbits around general Kerr-Newman black holes in three tables. The radial stability and rotation directions of the radii are also discussed.

 (3) For photon orbits between the above two planes($0<v<1$), quantitative and qualitative results are also provided. A critical inclination  angle, $v_{cr}=\frac{-1 + 4 u}{u (u+6)}$, is found for extremal Kerr-Newman black hole case. If $v<v_{cr}$, there is only one spherical photon orbit outside the event horizon for a rapidly rotating black hole ($u>\frac{1}{4}$). If $v>v_{cr}$, there are two orbits. If $v=v_{cr}$, there is only one orbit whose radii is analytically given. While for a slowly rotating black hole ($u<\frac{1}{4}$), there are always two spherical photon orbits outside the event horizon for any inclination angle. For photon orbits around nonextremal Kerr-Newman black hole,  we show a critical surface in the ($u,w,v$) parameter space which separates the regions with four and two positive solutions of the most general sextic equation.

Here we mainly focus on the radii of the spherical photon orbits around a Kerr-Newman black hole and the number of the orbits outside the black hole event horizon. 
We ignore the angular motion of the photon. It will be interesting to take it into account and to explore the angular motion of photon on spherical photon orbits. 
One can also study the generic light trajectories and light bending around a Kerr-Newman black hole.

\begin{acknowledgements}
This work is partially supported by Guangdong Major Project of Basic and Applied Basic Research (No. 2020B0301030008), Science and Technology Program of Guangzhou (No. 2019050001) and Natural Science Foundation of Guangdong Province (No. 2020A1515010388, No. 2020A1515010794).
\end{acknowledgements}


\begin{thebibliography}{99}
\bibitem{LIGOScientific:2016aoc}
B.~P.~Abbott \textit{et al.} [LIGO Scientific and Virgo],
``Observation of Gravitational Waves from a Binary Black Hole Merger,''
Phys. Rev. Lett. \textbf{116}, no.6, 061102 (2016).
\bibitem{LIGOScientific:2020iuh}
R.~Abbott \textit{et al.} [LIGO Scientific and Virgo],
``GW190521: A Binary Black Hole Merger with a Total Mass of $150  M_{\odot}$,''
Phys. Rev. Lett. \textbf{125}, no.10, 101102 (2020).
\bibitem{LIGOScientific:2020ufj}
R.~Abbott \textit{et al.} [LIGO Scientific and Virgo],
``Properties and Astrophysical Implications of the 150 M$_\odot$ Binary Black Hole Merger GW190521,''
Astrophys. J. Lett. \textbf{900}, no.1, L13 (2020).
\bibitem{LIGOScientific:2020zkf}
R.~Abbott \textit{et al.} [LIGO Scientific and Virgo],
``GW190814: Gravitational Waves from the Coalescence of a 23 Solar Mass Black Hole with a 2.6 Solar Mass Compact Object,''
Astrophys. J. Lett. \textbf{896}, no.2, L44 (2020).


\bibitem{EventHorizonTelescope:2019dse}
K.~Akiyama \textit{et al.} [Event Horizon Telescope],
Astrophys. J. Lett. \textbf{875}, L1 (2019).
\bibitem{EventHorizonTelescope:2019pgp}
K.~Akiyama \textit{et al.} [Event Horizon Telescope],
Astrophys. J. Lett. \textbf{875}, no.1, L5 (2019).
\bibitem{EventHorizonTelescope:2019ggy}
K.~Akiyama \textit{et al.} [Event Horizon Telescope],
Astrophys. J. Lett. \textbf{875}, no.1, L6 (2019).

\bibitem{EventHorizonTelescope:2022xnr}
K.~Akiyama \textit{et al.} [Event Horizon Telescope],
Astrophys. J. Lett. \textbf{930}, no.2, L12 (2022).


\bibitem{Cunha:2018acu}
P.~V.~P.~Cunha and C.~A.~R.~Herdeiro,
Gen. Rel. Grav. \textbf{50}, no.4, 42 (2018).
\bibitem{Cunha:2017eoe}
P.~V.~P.~Cunha, C.~A.~R.~Herdeiro and E.~Radu,
Phys. Rev. D \textbf{96}, no.2, 024039 (2017).
\bibitem{Gralla:2019xty}
S.~E.~Gralla, D.~E.~Holz and R.~M.~Wald,
Phys. Rev. D \textbf{100}, no.2, 024018 (2019).
\bibitem{Cvetic:2016bxi}
M.~Cvetic, G.~W.~Gibbons and C.~N.~Pope,
Phys. Rev. D \textbf{94}, no.10, 106005 (2016).

\bibitem{Lu:2019zxb}
H.~Lu and H.~D.~Lyu,
Phys. Rev. D \textbf{101}, no.4, 044059 (2020).
\bibitem{Ma:2019ybz}
L.~Ma and H.~Lu,
Phys. Lett. B \textbf{807}, 135535 (2020).
\bibitem{Yang:2019zcn}
R.~Q.~Yang and H.~Lu,
Eur. Phys. J. C \textbf{80}, no.10, 949 (2020).
\bibitem{Liu:2019rib}
H.~S.~Liu, Z.~F.~Mai, Y.~Z.~Li and H.~L\"u,
Sci. China Phys. Mech. Astron. \textbf{63}, 240411 (2020).
\bibitem{Hod:2020pim}
S.~Hod,
Phys. Rev. D \textbf{101}, no.8, 084033 (2020).
\bibitem{Hod:2013jhd}
S.~Hod,
Phys. Lett. B \textbf{727}, 345-348 (2013).


\bibitem{Cardoso:2016rao}
V.~Cardoso, E.~Franzin and P.~Pani,
Phys. Rev. Lett. \textbf{116}, 171101 (2016).
[erratum: Phys. Rev. Lett. \textbf{117}, 089902 (2016)]

\bibitem{Darwin1958}
C. G. Darwin,
Proc. R. Soc. Lond. A \textbf{249}, 180 (1959).
\bibitem{Darwin1961}
C. G. Darwin,
Proc. R. Soc. Lond. A \textbf{263}, 39 (1961).
\bibitem{Virbhadra:1999nm}
K.~S.~Virbhadra and G.~F.~R.~Ellis,
Phys. Rev. D \textbf{62}, 084003 (2000).
\bibitem{Claudel:2000yi}
C.~M.~Claudel, K.~S.~Virbhadra and G.~F.~R.~Ellis,
J. Math. Phys. \textbf{42}, 818-838 (2001).

\bibitem{Bardeen:1972fi}
J.~M.~Bardeen, W.~H.~Press and S.~A.~Teukolsky,
Astrophys. J. \textbf{178}, 347 (1972).
\bibitem{Teo2003}
E. Teo,
Gen. Relativ. Gravit. \textbf{35}, 1909-1926 (2003).

\bibitem{Cunha:2020azh}
P.~V.~P.~Cunha and C.~A.~R.~Herdeiro,
Phys. Rev. Lett. \textbf{124}, no.18, 181101 (2020).
\bibitem{Guo:2020qwk}
M.~Guo and S.~Gao,
Phys. Rev. D \textbf{103}, no.10, 104031 (2021).
\bibitem{Wei:2020rbh}
S.~W.~Wei,
Phys. Rev. D \textbf{102}, no.6, 064039 (2020).
\bibitem{Ghosh:2021txu}
R.~Ghosh and S.~Sarkar,
Phys. Rev. D \textbf{104}, no.4, 044019 (2021).



\bibitem{Yang:2012he}
H.~Yang, D.~A.~Nichols, F.~Zhang, A.~Zimmerman, Z.~Zhang and Y.~Chen,
Phys. Rev. D \textbf{86}, 104006 (2012).
\bibitem{Hod:2012ax}
S.~Hod,
Phys. Lett. B \textbf{718}, 1552-1556 (2013).
\bibitem{Tavlayan:2020cso}
A.~Tavlayan and B.~Tekin,
Phys. Rev. D \textbf{102}, no.10, 104036 (2020).



\bibitem{Hsiao:2019ohy}
Y.~W.~Hsiao, D.~S.~Lee and C.~Y.~Lin,
Phys. Rev. D \textbf{101}, no.6, 064070 (2020).
\bibitem{Hsieh:2021scb}
T.~Hsieh, D.~S.~Lee and C.~Y.~Lin,
Phys. Rev. D \textbf{103}, no.10, 104063 (2021).

\bibitem{Carter:1968rr}
B.~Carter,
Phys. Rev. \textbf{174}, 1559-1571 (1968).

\bibitem{Wang:2022ouq}
C.~Y.~Wang, D.~S.~Lee and C.~Y.~Lin,
[arXiv:2208.11906].

\end{thebibliography}
\end{document}